\documentclass{jpsj-suppl}
\usepackage{txfonts} %Please comment out this line unless the txfonts package is availabe in your LaTeX system.

\newcommand{\lsim}
 {\ \raise.35ex\hbox{$<$}\kern-0.75em\lower.5ex\hbox{$\sim$}\ }
\newcommand{\gsim}
 {\ \raise.35ex\hbox{$>$}\kern-0.75em\lower.5ex\hbox{$\sim$}\ }
\def\journal #1#2#3#4{#1 {\bf #2} (#4) #3}
\def\PR{Phys.\ Rev.}
\def\PRB{Phys.\ Rev.\ B}
\def\PRL{Phys.\ Rev.\ Lett.}

\def\JPCM{J.\ Phys.\ Cond.\ Mat.}

\def\JPSJ{J.\ Phys.\ Soc.\ Jpn.}

\def\PP{Phys.\ Proc.}
\title{Variational Approach to Localization Length for 
Two-Dimensional Hubbard Model}

\author{Shun \textsc{Tamura} and Hisatoshi \textsc{Yokoyama}}

\inst{Department of physics, Tohoku University, Sendai 
980-8578, Japan}

\email{shun@cmpt.phys.tohoku.ac.jp}

\recdate{September 30, 2013}

\abst{As a measure to ascertain whether a system is metallic or insulating, 
localization length $\lambda_N$, which represents the spread of electron 
distribution, can be a useful quantity, especially for approaching a 
metal-insulator transition from the insulator side. 
We try to calculate $\lambda_N$ using a variational Monte Carlo method 
for normal (paramagnetic), superconducting and antiferromagnetic states 
in the square-lattice Hubbard model. 
It is found that the behavior of $\lambda_N$ is consistent with what is 
expected from other quantities, and gives information complementary to 
another measure, the Drude weight. }

\kword{metal-insulator transition, localization length, Drude weight, 
variational Monte Carlo method, Hubbard model}

\begin{document}
\maketitle

\section{Introduction}
Because, in many strongly correlated electron systems, itinerancy and 
localization of conduction electrons are vital to their properties, 
appropriate and convenient measures to determine whether the system is 
a metal or an insulator have been pursued for long years. 
Various physical quantities are available to consider metal-insulator 
transitions, such as quasi-particle renormalization factor $Z$, charge 
structure factor $N({\bf q})$, chemical potential $\mu$ and the Drude weight 
$D$, DC component of conductivity, introduced by Kohn\cite{Kohn,Scalapino}.
Because these quantities are obtained within ground states, the nature 
of being a metal or an insulator is already inherent in the ground-state 
wave functions. 
Correctly, $N({\bf q})$ and $\mu$ are not measures, and the vanishing 
of $Z$ do not necessarily indicate 
an insulator, but represent a gap opening in some degree of freedom. 
Actually, $Z$ do not distinguish an $s$-wave superconducting (SC) state 
(gapped in the spin sector, but gapless in the charge sector) from an 
insulating state (gapped in the charge sector). 
As for $\mu$ \cite{chemicalP}, we have to find a jump as a function of 
electron density $n$ ($=N/N_{\rm s}$; $N$: electron number, $N_{\rm s}$: 
site number) or doping rate $\delta$ ($=1-n$), in addition to differentiating 
the total energy with respect to $n$. 
It is a little laborious. 
Here, we are interested in strongly correlated systems, for which a 
variational Monte Carlo (VMC) method\cite{VMC} is very useful for its exact 
treatment of local electron correlation without a minus sign problem.
Thus, in this work, we discuss useful measures of metal-insulator transitions 
for VMC calculations, namely, $D$ and in particular localzation length 
$\lambda_N$. 
We adopt a Hubbard model on the square lattice, which is a plausible model 
of the cuprate superconductors, and undergoes a Mott transition at half 
filling at $U\sim W$ (band width), unless an antiferromagnetic 
(AF) order is assumed. 
\par

Since the discovery of cuprate superconductors, $D$ has been intensively 
studied, because $D$ is directly obtained from experiments; in particular, 
anomalous $\delta$ dependence of superfluid density 
$\rho_{\rm s}$\cite{Uemura}, which is equivalent to $D$ in SC 
states\cite{Scalapino}, 
in cuprates is regarded as indubitable evidence of a doped Mott insulator. 
To obtain $D$ in the two-dimensional Hubbard model, we have to calculate 
the ground-state energy of the Hamiltonian\cite{Drude_VMC}, 
\begin{eqnarray}
{\cal H}(\bold{A})=-t\sum_{<i,j>,\:\sigma}\left[e^{i\bold{A}\cdot
(\bold{r}_i-\bold{r}_j)}c_{i\sigma}^\dag c_{j\sigma}+\mathrm{H.c.}\right]
+U\sum_j n_{j\uparrow}n_{j\downarrow},
\label{eq:Hamiltonian}
\end{eqnarray}
where $c_{j\sigma}(c_{j\sigma}^\dag)$ is the electron annihiration (creation) 
operator of spin $\sigma$ at site $j$, 
$n_{j\sigma}=c_{j\sigma}^\dag c_{j\sigma}$, 
$t$ (hopping integral) and $U$ (onsite interaction) are positive variables, 
and $\bold{A}$ is a virtual vector potential. 
$D$ is given by $D=d^2E(A)/dA^2$ with 
$E(A)=\langle{\cal H}(\bold{A})\rangle$. 
If the system is metallic (insulating), $D$ is finite (vanishes). 
In variation theories, however, a metal-insulator transition had not been 
described by means of $D$ until recently\cite{Millis}, namely, $D$ remained 
positive finite in insulating states even if binding factors between a doubly 
occupied site (doublon) and an empty site (holon)\cite{D-H} are introduced, 
with which Mott transitions are definitely described in terms of other 
quantities. 
Recently, this problem was solved as far as the range of $U\gtrsim U_{\rm c}$ 
($U_{\rm c}/t$: Mott transition point) is concerned\cite{Drude_VMC}. 
The ground state of eq.~(\ref{eq:Hamiltonian}) must be essentially complex, 
because the matrix elements are complex owing to the Peierls phase 
$e^{i\bold{A}\cdot(\bold{r}_i-\bold{r}_j)}$.
Therefore, an appropriate wave function should have a configuration-dependent 
phase factor ${\cal P}_\theta$\cite{Drude_VMC}. 
Thereby, not only Mott transitions are clearly identified by $D$, but linear 
behavior of $\rho_{\rm s}(\delta)$ (Uemura plot)\cite{Uemura} is obtained. 
We now know that this type of phase factors are indispensable for addressing 
current-carrying states such as various flux states\cite{flux} for 
intermediate and strong correlations. 
On the other hand, for insulators in weakly correlated regimes such as 
Slater-type AF insulators, we found that appropriate trial 
wave functions for finite {\bf A} should consist of multiple determinants; 
it seems technically difficult to treat them with the present VMC 
scheme\cite{Drude_VMC}. 
In calculating $D$, fine tuning of the trial functions seems necessary 
according to individual cases. 
\par

Although $D$ gives useful information in a metallic regime, $D$ does not 
tell how rigidly electrons are bound in an insulating regime, where 
$D$ always vanishes.  
As an alternative measure of metal-insulator transition from the insulator 
side, Resta\cite{Resta,R-S} introduced localization length $\lambda_N$, 
which approaches the spread of the electronic distribution, 
$\sqrt{[\langle(\sum_j x_j)^2\rangle-\langle\sum_j x_j\rangle^2]/N}$, 
as the system size increases. 
$\lambda_N$ represents how electrons can broaden in an insulating state; 
we can judge that the system is insulating (metallic), if $\lambda_N$ 
remains finite (diverges) as the system size is increased to infinity. 
Thus, a Mott transition point can be determined by the diverging point of 
$\lambda_N$, without carrying out differentiating operations in contrast 
to $D$. 
In early studies for one-dimensional systems, $\lambda_N$ or a corresponding 
susceptibility was calculated using exact diagonalization\cite{R-S}, 
quantum Monte Carlo method\cite{QMC}, and density matrix renormalization 
group\cite{DMRG}.   
Regarding VMC, $\lambda_N$ was calculated for a hydrogen chain\cite{H_chain}; 
it seems that $\lambda_N$ can be a good measure.
\par

In the following, we calculate the localization length in a Hubbard model 
on the square lattice [eq.~(\ref{eq:Hamiltonian}) with ${\bf A}={\bf 0}$]
using a VMC method, to confirm that $\lambda_N$ is an appropriate measure of 
a Mott transition. 
In sec.~\ref{sec:method}, we describe the method used in this study. 
In sec.~\ref{sec:results}, we show results of $\lambda_N$, and have 
discussions.  
A conclusion is given in sec.~\ref{sec:summary}. 
\par 

%---------------------
\section{Method\label{sec:method}}
In this section, we briefly explain variational wave functions used 
and the localization length $\lambda_N$. 
We study $\lambda_N$ with three types of wave functions: 
(i) a $d_{x^2-y^2}$-wave superconducting (SC) or singlet-pairing 
state $\Psi_\mathrm{SC}$, 
(ii) a paramagnetic or normal state $\Psi_\mathrm{N}$, and 
(iii) an AF state $\Psi_\mathrm{AF}$. 
Because these functions are applied to a highly correlated Hamiltonian, 
we adopt short-range Jastrow-type wave functions, which are composed of 
two parts: $\Psi={\cal P}\Phi$. 
Here, $\Phi$ represents a one-body wave function, $\Phi=\Phi_\mathrm{SC}$, 
$\Phi_\mathrm{N}$ or $\Phi_\mathrm{AF}$, each of which is a solution of 
the Hartree-Fock approximation without imposing the SCF condition. 
They are explicitly written as, 
\begin{eqnarray}
\Phi_\mathrm{SC}&=&\left(\sum_\bold{k}a_\bold{k}c_{\bold{k}\uparrow}^\dag
c_{-\bold{k}\downarrow}^\dag\right)^{N/2}|0\rangle, 
%\hspace{5mm}\mathrm{with}\hspace{5mm}
\qquad
a_\bold{k}=\frac{\Delta_d(\bold{k})}{\varepsilon_\bold{k}-\zeta
+\sqrt{(\varepsilon_\bold{k}-\zeta)^2+|\Delta_d(\bold{k})|^2}}, \\
\Phi_\mathrm{N}&=&
\prod_{k\in\mathrm{FS},\ \sigma}c_{k\sigma}^\dag|0\rangle,
\qquad\qquad\mbox{(FS: Fermi sea)} \\
\Phi_\mathrm{AF}&=&\prod_{\bold{k},\sigma}\left(\alpha_\bold{k}
c_{\bold{k}\sigma}^\dag+\tilde\sigma\beta_{\bf k}c_{\bold{k+Q}\sigma}^\dag
\right)|0\rangle,
%\hspace{5mm}\mathrm{with}\hspace{5mm}
\qquad
\alpha_\bold{k}\left(\beta_\bold{k}\right)=\sqrt{\frac{1}{2}\left(
1-(+)\frac{\varepsilon_\bold{k}}{\sqrt{\varepsilon_\bold{k}^2+
\Delta_\mathrm{AF}^2}}
\right)}.
\end{eqnarray}
In $d$-wave BCS wave function $\Phi_\mathrm{SC}$, 
$\varepsilon_\bold{k}=-2t(\cos k_x+\cos k_y)$, $\zeta$ is a variational 
parameter corresponding to the chemical potential in the limit of 
$U/t\rightarrow 0$, and the pair potential is assumed as 
$\Delta_d(\bold{k})=\Delta(\cos k_x-\cos k_y)$ with $\Delta$ being a 
variational parameter. 
In $\Phi_\mathrm{AF}$, $\tilde\sigma=\pm 1$ according to $\sigma=\uparrow$ 
or $\downarrow$, $\bold{Q}=(\pi,\pi)$, and $\Delta_\mathrm{AF}$ is 
a variational parameter related to the staggered moment.
As for the many-body part ${\cal P}$ in $\Psi$, we consider only dominant 
factors, ${\cal P}={\cal P}_\mathrm{G}{\cal P}_Q$, in common for the three
$\Psi$'s. 
The most fundamental one is the onsite (Gutzwiller) factor, 
${\cal P}_\mathrm{G}=\prod_j[1-(1-g)n_{j\uparrow}n_{j\downarrow}]$ 
with $g$ being a variational parameter ($0\leq g\leq 1$). 
The second factor is a doublon-holon binding factor between nearest-neighbor 
(NN) sites \cite{D-H}, which is indispensable for inducing a Mott 
transition: 
${\cal P}_Q=\prod_j[1-\mu_1Q_j^1]\prod_i[1-\mu_2Q_i^2]$, with
$Q_j^1=n_{j\uparrow}n_{j\downarrow}\prod_\tau 
[1-(1-n_{j+\tau\uparrow})(1-n_{j+\tau\downarrow})]$, and
$Q_j^2=(1-n_{j\uparrow})(1-n_{j\downarrow})\prod_\tau
(1-n_{j+\tau\uparrow}n_{j+\tau\downarrow})$. 
Here, $\mu_1$ and $\mu_2$ are variational parameters 
($0\leq \mu_1,\:\mu_2\leq 1$), and $\tau$ runs over the NN sites of site $j$.
Some properties of these wave functions when applied to 
eq.~(\ref{eq:Hamiltonian}) are known \cite{YTOT,YOT,YOTKT,Drude_VMC}.
At half filling, it was confirmed through the behavior of $Z$, $N({\bf q})$, 
$D$, etc. that $\Psi_\mathrm{SC}$ and $\Psi_\mathrm{N}$ exhibit first-order 
Mott transitions at $U/t\sim 6.5$ and $8.5$, respectively. 
On the other hand, it is expected that an AF state becomes the ground state 
and insulating for any $U/t$ $(>0)$ without transitions. 
Actually, $\Psi_\mathrm{AF}$ is insulating and has the lowest energy among 
the three $\Psi$'s for $U/t\gtrsim 1.5$. 
For $0<U/t\lesssim1.5$, the AF order seems to vanish in $\Psi_\mathrm{AF}$, 
but this is irrelevant to the following discussions. 
In less-than-half-filled cases, the three $\Psi$'s are always metallic. 
\par

In an insulator (metal), the ground-state wave function is localized 
(delocalized). 
In a system under the periodic boundary condition, localization can be 
treated in parallel with the modern theory of polarization\cite{Resta,R-S}. 
The degree of localization is embedded in a complex number $z_N$ defined 
below, whose modulus is the definition of localization length $\lambda_N$:  
\begin{eqnarray}
\lambda_N=
\left(\frac{L}{2\pi}\right)\sqrt{-\frac{\ln|z_N|^2}{N}} 
\hspace{5mm}\mathrm{with}\hspace{5mm}
z_N=\langle\Psi|e^{i\frac{2\pi}{L}X}|\Psi\rangle
\qquad (L\rightarrow\infty),
\label{eq:lambda}
\end{eqnarray}
where  $L$ is the system size in one ($x$) direction, and $X$ is the sum 
of $x$ coordinates of electrons: $X=\sum_{j,\sigma}x_jn_{j\sigma}$.
If we expand $z_N$ with respect to $2\pi/L$ and substitute it in the 
expression of $\lambda_N$ in eq.~(\ref{eq:lambda}), the definition of 
localization length, 
$\sqrt{(\langle X^2\rangle-\langle X\rangle^2)/N}$, is obtained as the leading 
term. 
Therefore, we should check $L$ dependence of $\lambda_N$ to monitor the 
deviation owing to higher-order terms. 
When the state is extremely localized like a $\delta$ function, $z_N$ 
becomes 1; $z_N$ is finite ($0<|z_N|<1$) for general localized states, 
whereas $z_N$ vanishes when the state is delocalized. 
Equivalently, $\lambda_N$ converges to a finite value for an insulator, 
but diverges as $L$ increases for a metallic system.
Using the above wave functions, the expectation value of $z_N$ is 
easily calculated by VMC. 
\par

We carry out a series of VMC calculations to estimate $z_N$ using 
correlated measurement with quasi-Newton algorithm for large systems 
of $L\times L$ sites ($L=12$-$20$). 
A typical sample number in this study is $M=2.4\times10^5$. 
\par

%---------------------
\section{Results\label{sec:results}}
\begin{figure} 
\begin{center}
\includegraphics[width=15cm]{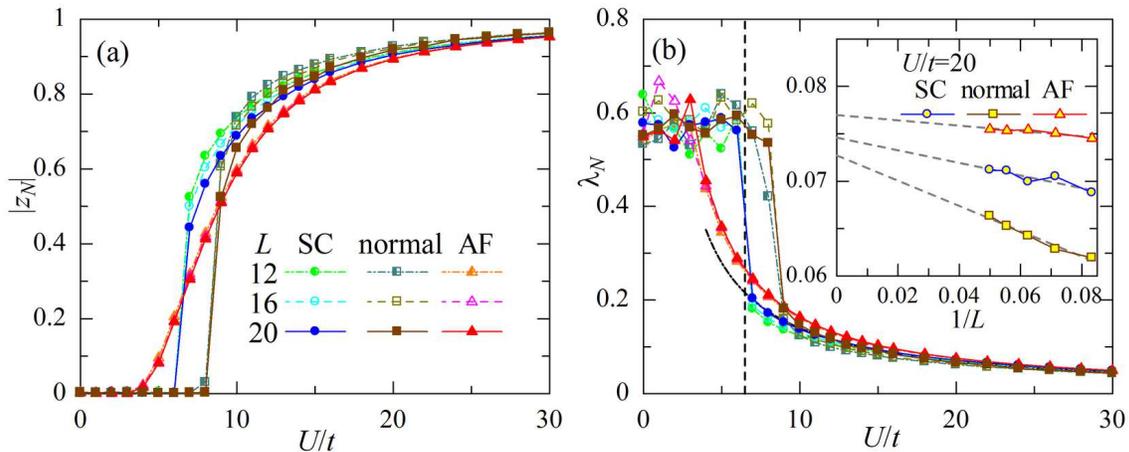}
\end{center}
\caption{(Color online) 
(a) Absolute values of $z_N$ are shown for the three $\Psi$'s as a function 
of $U/t$ at half filling for three $L$'s.
(b) Localization length are compared among the three wave functions. 
The symbols are common to (a). 
A dash-dotted curve of $\propto t/U$ is added as a guide for eyes. 
The vertical dashed line indicates the metal-insulator transition point of 
$\Psi_{\rm SC}$.
Inset in (b) shows the extrapolation of $\lambda_N$ to $L\rightarrow\infty$ 
for the three states for $U/t=20$ from the data for $L=12$-$20$.
The number of samples is $2.4\times10^5$.
}
\label{fig:f1}
\end{figure}
Let us start with $U/t$ dependence of $|z_N|$ at half filling. 
In Fig. \ref{fig:f1}(a), we plot $|z_N|$ obtained by VMC calculations 
for the three states. 
For $\Psi_{\rm SC}$, $|z_N|$ is almost zero for $U/t\leq6$, but 
suddenly increases with a jump at approximately $U/t=6.5$, and tends to 
unity for $U/t\geq7$. 
As mentioned above, this behavior of $|z_N|$ indicates that the system is 
conductive (insulating) for $U/t<6.5$ ($U/t>6.5$), and at $U/t=6.5$ a sharp 
(first-order-like) SC-insulator transition arises. 
This result is quantitatively consistent with that of the previous 
studies\cite{YTOT,YOT,Drude_VMC}, in which this Mott transition was 
confirmed by various quantities such as a SC pairing correlation function
and the Drude weight. 
The behavior of $|z_N|$ of $\Psi_{\rm N}$ is similar to that of 
$\Psi_{\rm SC}$ except for the Mott transition point, and is consistent 
with that of the previous results. 
Although appreciable system-size dependence in $|z_N|$ is observed both for 
$\Psi_{\rm SC}$ and $\Psi_{\rm N}$, especially near the transition points, 
they will not qualitatively change the features of Mott transition. 
We will return to this point again in connection with $\lambda_N$. 
$|z_N|$ of $\Psi_{\rm AF}$ is again almost zero for $0<U/t\leq3$, but 
this time $|z_N|$ increases smoothly for $U/t>3$, as $U/t$ increases, 
without anomalous behavior like a jump. 
Thus, $\Psi_{\rm AF}$ is insulating at least for $U/t>3$. 
Considering $\Psi_{\rm AF}$ is insulating for more smaller values of $U/t$, 
we expect that the value of $|z_N|$ is very small but still non zero 
for $U/t<3$. 
In this range, it is difficult to determine the accurate behavior of 
$|z_N|$, because statistical errors in VMC exceed the magnitude of $|z_N|$. 
We will discuss this point later again. 
The system-size dependence for $\Psi_{\rm AF}$ is by far smaller than 
those for $\Psi_{\rm SC}$ and $\Psi_{\rm N}$. 
\par

We turn to the localization length $\lambda_N$, which is calculated 
from $z_N$ through eq.~(\ref{eq:lambda}). 
In Fig.~\ref{fig:f1}(b), we plot $\lambda_N$ obtained from the data in 
Fig.~\ref{fig:f1}(a). 
As a whole, $\lambda_N$ tends to decrease as $U/t$ increases; this behavior 
agrees with our intuition that the electrons are more localized as the 
repulsive interaction becomes stronger. 
Corresponding to the behavior of $|z_N|$, $\lambda_N$ for $\Psi_{\rm SC}$ 
and $\Psi_{\rm N}$ exhibit discontinuities at $U/t=6.5$ and $8.5$, 
respectively; $\lambda_N$ of $\Psi_{\rm AF}$ is smooth for $U/t\gtrsim 3$. 
\par

First, we discuss the insulating regimes.
Here, each $\lambda_N$ decreases approximately as a 
function of $t/U$ ($=J/4t$), indicating that the insulating regime is 
effectively described by an AF Heisenberg model with the exchange coupling 
$\sim J$, which is the sole energy scale. 
Now, we look at the system-size dependence.  
In the inset of Fig.~\ref{fig:f1}, we plot $\lambda_N$ at $U/t=20$ in the 
insulating regime as a function of $1/L$. 
The system-size dependence is the largest in $\Psi_{\rm N}$ and the smallest 
in $\Psi_{\rm AF}$, but the order among the three does not seem to change 
in the limit of $L\rightarrow\infty$. 
The localization length of $\Psi_{\rm SC}$ is a little longer than that 
of $\Psi_{\rm N}$, indicating electrons in SC is more mobile than in the 
normal state even in a insulating regime. 
This corresponds to the fact that the kinetic energy is lower for 
$\Psi_{\rm SC}$ than for $\Psi_{\rm N}$ (not shown), namely, 
a kinetic-energy-driven SC ($\Psi_{\rm N}\rightarrow\Psi_{\rm SC}$) 
transition is realized, as previous studies showed\cite{YTOT}. 
Incidentally, for smaller values of $U/t$ ($U_{\rm c}<U<20t$), we cannot 
implement reliable extrapolations for $\Psi_{\rm SC}$ and especially 
$\Psi_{\rm N}$, because the system-size dependence of $\lambda_N$ becomes 
somewhat irregular. 
It seems that this is not only owing to the sampling errors but to the 
short range nature of the trial wave functions. 
It is possible that a long-range correlation factor is necessary for systems 
of large $\lambda_N$ to stably estimate $\lambda_N$ near Mott transitions. 
The notion of a kinetic-energy-driven transition is similarly applicable 
to the AF transition ($\Psi_{\rm N}\rightarrow\Psi_{\rm AF}$), because 
$\lambda_N^{\rm AF}>\lambda_N^{\rm N}$. 
The reason $\lambda_N^{\rm AF}$ is the largest is that electrons are more 
likely to hop to NN sites owing to the spin alternate configurations in 
$\Psi_{\rm AF}$. 
\par

\begin{figure}
\begin{center}
\includegraphics[width=15cm]{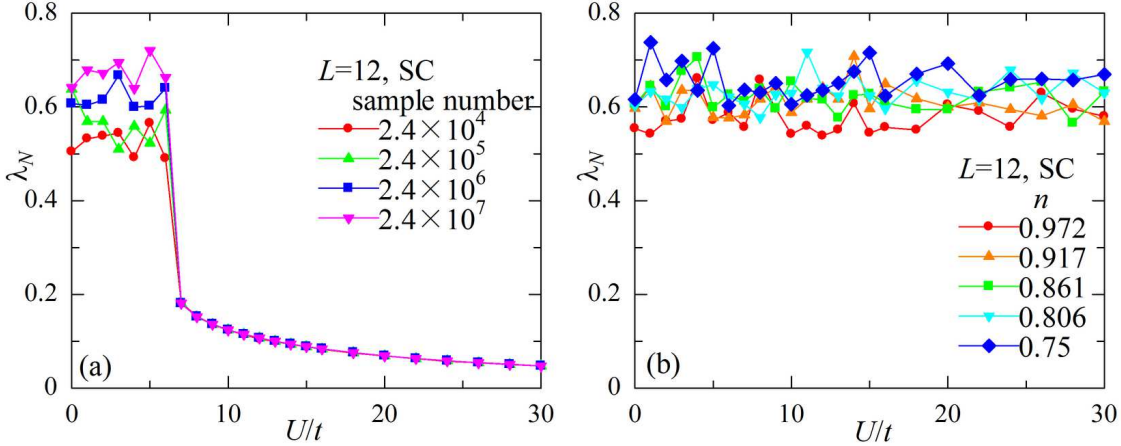}
\end{center}
\caption{(Color online)
(a) 
Expectation values of localization length at half filling are compared for 
different sample numbers in the VMC calculations for $\Psi_{\rm SC}$.
(b) 
Localization length estimated with the same VMC procedure for several 
doped systems. 
$n$ indicates an electron density.
The number of sample is $2.4\times10^5$.
}
\label{fig:f2}
\end{figure}

Next, we consider the results of $\lambda_N$ in weakly correlated regimes. 
In the insulating regimes discussed above, the expectation values of 
$\lambda_N$ converges at finite values, as expected.  
On the other hand, in weakly correlated regimes ($U<U_{\rm c}$ for 
$\Psi_{\rm SC}$ and $\Psi_{\rm N}$; $U\lesssim 3t$ for $\Psi_{\rm AF}$), 
$\lambda_N$ does not diverge as $L\rightarrow\infty$, but has scattered 
values around 0.6 irrespective of the system size, as found in 
Fig.~\ref{fig:f1}(b). 
An origin of this result probably lies in the statistical errors in VMC 
sampling. 
The expectation values of $|z_N|$ for these correlation strengths are 
extremely close to zero but positive finite. 
Therefore, if the numerical error $\epsilon$ exceeds the correct expectation 
value, the estimation by VMC yields an arbitrary value of order $\epsilon$ 
as $|z_N|$. 
To verify this possibility, we carry out a series of VMC calculations for 
$\Psi_{\rm SC}$ by widely varing the sampling number $M$ for a fixed $L$ 
($=12$); the statistical errors are proportional to $1/\sqrt{M}$. 
The results are plotted in Fig.~\ref{fig:f2}(a). 
We find that the value of $\lambda_N$ steadily increases as $M$ increases, 
although the increments are small because $\lambda_N$ is a logarithmic 
function of $|z_N|$ [eq.~(\ref{eq:lambda})]. 
Thus, in order to estimate $\lambda_N$ accurately for a system with large 
$\lambda_N$, it is necessary to precisely determine $|z_N|$ of a large 
negative power. 
In this connection, we calculate $\lambda_N$ for less-than-half-filled 
or doped systems, which are metallic for any value of $U/t$, with the  
same sample number as in Fig.~\ref{fig:f1}(b).  
As shown in Fig.~\ref{fig:f2}(b), the results are scattered around 
$\lambda_N=0.6$ in broad accordance with that for the weakly interacted
half-filled case in Fig.~\ref{fig:f1}(b). 
We can perceive a slight tendency for $\lambda_N$ to increases as 
electron density ($n$) decreases.  
Anyway, the present VMC calculations yield a reliable value of $\lambda_N$ 
for an insulator of $\lambda_N\lesssim 0.5$. 
\par

Finally, we touch on $\lambda_N$ of $\Psi_{\rm AF}$. 
Judging from other quantities, $\Psi_{\rm AF}$ is insulating at least 
for $U/t\gtrsim 1.5$, as mentioned. 
Therefore, $\lambda_N$ should converge at finite values in this range.  
Correspondingly, $\lambda_N$ in Fig.~\ref{fig:f1}(b) exhibits a smooth 
converged curve for large values of $U/t$ down to $U/t\sim 3$. 
However, at $U/t=3$, $\lambda_N$ reaches 0.6 still in the insulating phase. 
Thus, to address an insulator in weak-correlation regimes, a formalism 
or an algorithm in a new line seems necessary. 
\par

%----------------------------------------
\section{Summary\label{sec:summary}}
Localization length $\lambda_N$, which is the variance in coordinates 
of electrons, is calculated for the two-dimensional Hubbard model to 
distinguish a metal from an insulator, using a variational Monte Carlo 
method. 
$\lambda_N$ thus obtained definitely indicates a Mott transition point 
($U_{\rm c}$) by a discontinuity for a normal or a $d$-wave pairing state, 
whose $U_{\rm c}$ is at a correlation strength broadly of the band width. 
On the other hand, we found that reliable estimation of $\lambda_N$ by VMC 
is not easy for a system of $\lambda_N$ beyond a threshold value determined 
by the numerical errors ($\sim 0.6$ in the present setting), regardless of 
being a metal or an insulator. 
Therefore, the present scheme is not necessarily suitable to address weakly 
correlated insulators such as a Slater-type antiferromagnetic insulator. 
Nevertheless, we may say that $\lambda_N$ is a useful quantity to discuss 
Mott transitions using various Monte Carlo methods. 
\par

%\appendix
%\section{}

\end{document}